\documentclass[12pt]{article}
\usepackage{fancyhdr,fancybox,amssymb,epsfig, amsmath, latexsym,enumerate, mathtools}
\usepackage[ruled,vlined]{algorithm2e}
\usepackage{xargs}

\newcommand{\R}{\ensuremath{\mathcal{R}}}

\newcommand{\s}{\ensuremath{\sigma}}

\newcommand{\pk}{\texttt{pk}}

\newcommand{\sk}{\texttt{sk}}

\newcommandx{\KeyGen}[2][1=PP,2=\ensuremath{\ell}]{\texttt{KeyGen}\ensuremath{(#1, #2)}}
\newcommandx{\Sign}[2][1=\sk, 2=M]{\texttt{Sign}\ensuremath{(#1, #2)}}
\newcommandx{\Verify}[3][1=\pk, 2=M, 3=\s]{\texttt{Verify}\ensuremath{(#1, #2, #3)}}
\newcommandx{\ConvertSK}[2][1=\sk, 2=\ensuremath{\rho}]{\texttt{ConvertSK}\ensuremath{(#1, #2)}}
\newcommandx{\ConvertPK}[2][1=\pk, 2=\ensuremath{\rho}]{\texttt{ConvertPK}\ensuremath{(#1, #2)}}
\newcommandx{\ConvertSig}[4][1=\pk, 2=M, 3=\s, 4=\ensuremath{\rho}]{\texttt{ConvertSig}\ensuremath{(#1, #2, #3, #4)}}
\newcommandx{\ChangeRep}[4][1=\pk, 2=M, 3=\s, 4=\ensuremath{\mu}]{\texttt{ChangeRep}\ensuremath{(#1, #2, #3, #4)}}

\newcommandx{\KeyGennp}{\texttt{KeyGen}}
\newcommandx{\Signnp}{\texttt{Sign} }
\newcommandx{\Verifynp}{\texttt{Verify}}
\newcommandx{\ConvertSKnp}{\texttt{ConvertSK}}
\newcommandx{\ConvertPKnp}{\texttt{ConvertPK}}
\newcommandx{\ConvertSignp}{\texttt{ConvertSig}}
\newcommandx{\ChangeRepnp}{\texttt{ChangeRep}}

\newcommand{\Z}{\ensuremath{\mathbb{Z}}}

\usepackage{tikz}
\usepackage{wrapfig}

\newcommand{\cR}{\ensuremath{\mathcal{R}}}

\newcommand{\bv}{\ensuremath{\mathbf{v}}}
\newcommand{\bw}{\ensuremath{\mathbf{w}}}

\newcommand{\bA}{\ensuremath{\mathbf{A}}}

\usepackage{optidef}

\usepackage{dirtytalk}
\usepackage{minted}

\usepackage{url}
\usepackage{setspace}

\usepackage[plainpages=false,pdfpagelabels]{hyperref} 
\usepackage[letterpaper,margin=1.25in]{geometry}   
\usepackage[toc,title,page]{appendix}              
\usepackage{tocloft}

\usepackage{natbib}

\usepackage{fancyhdr,fancybox,amssymb,epsfig, amsmath, latexsym,enumerate}
\usepackage[ruled,vlined]{algorithm2e}
\usepackage{amsmath,amscd}
\usepackage{tikz}
\usetikzlibrary{shapes,arrows}
\usepackage{multirow}

\newtheorem{definition}{Definition}[section]
\newtheorem{theorem}{Theorem}[section]

\usepackage[utf8]{inputenc}
\usepackage{mathrsfs}
\usepackage{color,soul}
\usepackage{tikz}
\usetikzlibrary{matrix}
\usetikzlibrary{cd}
\usepackage[ruled,vlined]{algorithm2e}
\usepackage{tikz-network}
\usepackage{todonotes}
\usepackage{float}
\floatstyle{plaintop}

\newcommand{\tx}[1]{\texttt{#1}}

\def\Z{\mathbb{Z}}
\def\R{\mathbb{R}}

\def\<{\langle}
\def\>{\rangle}

\usepackage{thm-restate}

{\setlength{\fboxsep}{15pt}\begin{Sbox}\begin{minipage}}%
{\end{minipage}\end{Sbox}\fbox{\TheSbox}}
\newcommand{\qed}{\hfill \ensuremath{\Box}}

\usepackage{array}

\newcommand{\norm}[1]{\left\lVert#1\right\rVert}
\newcommand{\inorm}[1]{\norm{#1}_{\infty}}

\renewcommand{\figurename}{Algorithm}

\usepackage{titlesec}
\titleformat*{\section}{\large\bfseries}
\titleformat*{\subsection}{\normalsize\bfseries}
\usepackage[all,cmtip]{xy}

\begin{document}

\begin{center}
    \textbf{Security Analysis of Integer Learning with Errors with \\ Rejection Sampling}

    \bigskip

    Kyle Yates$^1$,
    Antsa Pierrottet$^1$,
    Abdullah Al Mamun$^2$,
    Ryann Cartor$^1$,\\
    Mashrur Chowdhury$^2$, and 
    Shuhong Gao$^1$

    \bigskip

    {\small $^1$School of Mathematical and Statistical Sciences, Clemson University,\\ Clemson, South Carolina, 29634

    \vspace{.05in}

$^2$Glenn Department of Civil Engineering, Clemson University,\\ Clemson, South Carolina, 29634}
\end{center}

\begin{abstract}

At ASIACRYPT 2018, a digital attack based on linear least squares was introduced for a variant of the learning with errors (\textsf{LWE}) problem which omits modular reduction known as the integer learning with errors problem (\textsf{ILWE}). In this paper, we present a theoretical and experimental study of the effectiveness of the attack when applied directly to small parameter \textsf{ILWE} instances found in popular digital signature schemes such as CRYSTALS-Dilithium which utilize rejection sampling. Unlike other studies which form \textsf{ILWE} instances based on additional information obtained from side-channel attacks, we
take a more direct approach to the problem
by constructing our \textsf{ILWE} instance from only the obtained signatures. We outline and introduce novel techniques in our simulation designs such as modular polynomial arithmetic via matrices in $\R$, as well as algorithms for handling large sample sizes efficiently. Our experimental results reinforce the proclaimed security of signature schemes based on \textsf{ILWE}. We additionally discuss the implications of our work and digital signatures as a whole in regards to real-world applications such as in Intelligent Transportation Systems (ITS).
\end{abstract}

\section{Introduction}

Digital signatures ensure the authenticity and validity of digital communications and interactions and are an essential component of any modern cryptographic protocol. In practice and implementation, algorithms currently used include EdDSA \cite{rfc8032}, RSA \cite{rfc8017}, and ECDSA \cite{rfc6979,NIST:SP800-186}, which are all standardized by the National National Institute of Standards and Technology (NIST) in the Federal Information Processing Standard (FIPS) 186-5 Digital Signature Standard (DSS) \cite{NIST:FIPS186-5}.

In recent years, quantum computing has become a major anticipated threat to currently used cryptographic systems (including digital signature schemes) due to quantum algorithms such as Shor's algorithm \cite{shororiginalpaper,shordetailedpaper} and Grover's algorithm \cite{groverorigpaper,groverexpandedpaper}. With this eventual threat, NIST has selected several post-quantum public key schemes and digital signature schemes for standardization which can be implemented on classical computers yet are resistant to attacks by quantum computers. Selected digital signature algorithms include CRYSTALS-Dilithium \cite{crystalsdilithium_ducas_2018} (standardized in FIPS 204 \cite{NIST:FIPS204}), SPHINCS+ \cite{SPHINCS+} (standardized in FIPS 205 \cite{NIST:FIPS205}), and FALCON \cite{FALCON} (FIPS upcoming).

One of the most popular areas of post-quantum cryptography is lattice-based cryptography, which CRYSTALS-Dilithium and FALCON are both based on. For $m \geq 1$ an integer, a subset $\Lambda \subseteq \mathbb{R}^m$ is a lattice if $\Lambda$ is a discrete additive subgroup of $\mathbb{R}^m$ and no points in $\Lambda$ are arbitrarily close to each other. Of particular interest in lattice-based cryptography is a class of lattices known as $q$-ary lattices. A lattice $\Lambda$ is a $q$-ary lattice if $q\Z^m \subseteq \Lambda \subset \Z^m$. Given an integer matrix $\bA$, a $q$-ary lattice may be defined more explicitly -- for instance, via
$$\Lambda = \mathcal{L}_q^\perp (\bA) = \left\{\bv \in \mathbb{Z}^m\ :\ \bA\bv \equiv \mathbf{0} \mod q \right\}.$$

Several computationally hard problems arise from lattices such as the shortest vector (\textsf{SVP}), $\gamma$-approximate shortest vector ($\gamma$-\textsf{SVP}),  $\gamma$-gap shortest vector ($\gamma$-\textsf{GapSVP}), closest vector (\textsf{CVP}), and bounded distance decoding (\textsf{BDD}) problems (see, e.g., \cite{c54b603f579b48a08b698bde47b71455} and the references within). These hard lattice problems (and several others) form the basic theoretical foundation for security in many lattice-based cryptographic schemes. In many of these schemes, more tangible extensions of these problems are used, such as the learning with errors (\textsf{LWE}) problem and the short integer solutions (\textsf{SIS}) problem. The basic idea of \textsf{LWE} involves solving noisy linear systems modulo some paramter $q$ and is at least as hard as some hard lattice problems, a result proved by Regev in 2005 \cite{Reg05}. Several popular variations of \textsf{LWE} exist and are used in practice such as module learning with errors (\textsf{MLWE}), which is described more in-depth alongside \textsf{SIS} in Section \ref{sec:LWE}.

In 2018, Bootle et al. \cite{bootle} introduced a digital attack based on linear least squares which is effective against an integer version of the \textsf{LWE} problem known as integer learning with errors (\textsf{ILWE}) in which elements do not undergo reduction modulo $q$. Although not always immediately apparent, \textsf{ILWE} is used in some constructions of lattice-based digital signature schemes when paired with the technique of rejection sampling (explained in Section \ref{section.rejection_sampling}). The attack in \cite{bootle} is applied to a lattice-based signature scheme known as BLISS \cite{cryptoeprint:2013/383} using information obtained via side-channel attacks to formulate an \textsf{ILWE} instance.

In the signing algorithm of CRYSTALS-Dilithium one step is based on the security of \textsf{ILWE} with rejection sampling. Previous other studies have looked at the effectiveness of this least squares attack against CRYSTALS-Dilithium (see, e.g., \cite{cryptoeprint:2019/715,cryptoeprint:2022/106, cryptoeprint:2023/896,cryptoeprint:2023/050,cryptoeprint:2024/1373,cryptoeprint:2025/820}). However, most of these attacks base themselves from information based on side-channel leakage. Furthermore, they do not look at the effectiveness of said attack for varying sizes of parameters. In this paper, we will look at a simpler but more direct approach to testing the attack from Bootle et al. against the \textsf{ILWE} structure prevalent in CRYSTALS-Dilithium without leaked information from side-channel attacks. This approach includes variations to parameters and underlying sampling distributions.

\subsection{Our Contributions}

We implement and apply simulations of the attack from Bootle et al. \cite{bootle} to rejection sampling procedures present in schemes such as CRYSTALS-Dilithium to gauge the effectiveness of the attack. Unlike the approach in \cite{bootle} and other aforementioned works which rely on information obtained via side-channel attacks, we opt for a more direct formulation of an \textsf{ILWE} instance using only the sample values obtained from the signing algorithm in CRYSTALS-Dilithium and similar schemes. In designing the implementation, we use novel techniques for transferring arithmetic in specific modules to matrix arithmetic over $\mathbb{R}$. We also introduce an algorithm to handle large sample sizes efficiently for the attacks from \cite{bootle}. Our experimental results highlight the importance of the underlying sampling distributions used in \cite{crystalsdilithium_ducas_2018} and help to highlight a range of parameters which are secure against these attacks without any assumptions of side-channel leakage that may occur.

\subsection{Organization of This Paper}

In Section \ref{section.preliminaries}, we outline the necessary notation and preliminaries including \textsf{LWE}, \textsf{SIS}, and the attack from Bootle et al. \cite{bootle}. In Section \ref{section.rejection_sampling}, we describe background on rejection sampling and relevant signature schemes which use rejection sampling such as BLISS and CRYSTALS-Dilithium. In Section \ref{section.implementation_strategies} we discuss our implementation strategies for matrix arithmetic and large sample sizes. Section \ref{section.concretesecurity} describes our simulation framework and experimental results. Section \ref{section.applications} discusses broader applications of our results, including applications in Intelligent Transportation Systems (ITS). Section \ref{section.conclusion} provides for concluding remarks.

\section{Preliminaries}\label{section.preliminaries}

\subsection{Notation }

Let $q$ be a positive integer. We denote $\Z_q = \Z \cap (-q/2,q/2]$ as the ring of centered representatives. Integers reduced modulo $q$ are always reduced into this ring. That is, for $\alpha\in \Z$, we have $\alpha' = \alpha \mod q$ as the unique element in $\Z_q$ such that $q$ divides $\alpha - \alpha'$. For $n$ a power of two, define the polynomial rings
\[
\mathcal{R} = \Z[X]/(X^n+1) \quad \text{and} \quad \mathcal{R}_q = \Z_q[X]/(X^n+1).
\]
For a polynomial $w = w_0  + w_1 X + \dots + w_{n-1}X^{n-1}\in \mathcal{R}$ (or $\mathcal{R}_q)$ or vector $w=(w_0,\dots,w_{n-1})\in \R^{n}$, we denote the standard 1, 2, and infinity norms as  

\[
\norm{w}_1 = \sum_{i=0}^{n-1} |w_i|, \quad \norm{w}_2 = \sqrt{\sum_{i=0}^{n-1} w_i^2}, \quad \inorm{w} = \max_{0\leq i < n}\{|w_i|\}
\]
respectively. For the vector $\bw = (w^{(0)}, \ldots , w^{(k-1)}) \in \cR^k$ (or $\mathcal{R}_q^k$), we similarly have 
\[
\norm{\textbf{w}}_1 = \sum_{j=0}^{k-1} \|w^{(j)}\|_1, \quad \norm{\textbf{w}}_2 = \sqrt{\sum_{j=0}^{k-1} \|w^{(j)}\|_2^2}, \quad \inorm{\textbf{w}} = \max_{0\leq j < k}\{\|w^{(j)}\|_{\infty}\}.
\]
For $v,w\in \mathcal{R}$ we define $\<v,w\> = v_0w_0 + v_1w_1 + \dots +v_{n-1}w_{n-1}$, where the $v_i$ and $w_i$'s are the respective coefficients of the polynomials. When $\textbf{v}=(v^{(0)},\dots , v^{(k-1)}),\textbf{w} = (w^{(0)},\dots, w^{(k-1)})\in \mathcal{R}^k$, we have $\< \textbf{v}, \textbf{w}\> = \< v^{(0)},w^{(0)}\> + \dots + \< v^{(k-1)}, w^{(k-1)}\>$.

We denote $\tx{coeff}(v) = (v_0,\dots,v_{n-1})$ as the coefficient vector of $v=v_0 + v_1 X + \dots+ v_{n-1} X^{n-1}\in \mathcal{R}$. When $\textbf{v}=(v^{(0)},\dots, v^{(k-1)})\in \mathcal{R}^k$, we use the notation $\tx{coeff}(\textbf{v}) = (\tx{coeff}(v^{(0)})|| \dots || \tx{coeff}(v^{(k-1)}))$. We also denote $\tx{wt}(v)$ as the Hamming weight of $v\in \mathcal{R}$, which is the number  of non-zero coefficients of $v$. When $\textbf{v}=(v^{(0)},\dots, v^{(k-1)})\in \mathcal{R}^k$, we use the notation $\tx{wt}(\textbf{v}) =\sum_{j=0}^{k-1} \tx{wt}(v^{(j)})$.

For $a\in \mathbb{R}$, we use $\lfloor a\rceil$ to denote the rounding of $a$ to the closest integer, rounding down in the case of a tie. For a vector  or polynomial $\textbf{a}$, we use $\lfloor \textbf{a}\rceil$ to denote the rounding of each entry or coefficient, respectively. 

For a vector $\textbf{a}$, we denote $\textbf{a}[i]$ as the $i$th entry of $\textbf{a}$ and $\textbf{a}[i\text{:}j]$ as the vector of length $j-i+1$ consisting of entries $i$ through $j$ of $\textbf{a}$. We extend this notation to a matrix $\textbf{A}$, letting $\textbf{A}[i,j]$ denote entry $(i,j)$ of \textbf{A} and $\textbf{A}[i\text{:}j,k\text{:}\ell]$ the submatrix of  \textbf{A} consisting of rows $i$ through $j$ and columns $k$ through $\ell$.

For a set $S$, we denote $a\xleftarrow{\$} S$ as the uniform random sampling of $a$ from the set $S$. For an arbitrary probability distribution $\chi$, we denote $a \leftarrow \chi$ as the sampling of $a$ from that distribution. We denote $\mathcal{D}_{\sigma}$ as the discrete Gaussian distribution over $\Z$ with standard deviation $\sigma$. Over $\Z$, the discrete Gaussian distribution $\mathcal{D}_{\sigma}$ assigns a probability proportional to $\exp (-\pi a^2/(\sigma/\sqrt{2\pi})^2)$ for each $a\in \Z$. An $n$-dimensional extension of $\mathcal{D}_{\sigma}$ to $\Z^n$ (or $\mathcal{R})$ may be constructed by sampling each entry (or coefficient) from $\mathcal{D}_{\sigma}$. We denote this $n$-dimensional extension as $\mathcal{D}_{\sigma}^n$.

\subsection{Short Integer Solution Problem}\label{sec:SIS}

We first introduce several popular problems in lattice-based cryptography which form the theoretical foundation for security. The first of which is the Short Integer Solution (SIS) problem, in which one must find a short integer combination of vectors equivalent to $0$ modulo some parameter $q$. We state the problem more formally in Definition \ref{def.SIS} below.

\begin{definition}[Short Integer Solution (SIS)]\label{def.SIS} 
Let $m, k, q$, and $\beta < q$ be positive integers. For a randomly-chosen matrix $\bA\leftarrow \Z_q^{m\times k}$, the \textup{\textsf{SIS}} problem is to find $\textup{\textbf{x}}\neq \textup{\textbf{0}} \in \Z^k$ such that $\bA \textup{\textbf{x}} \equiv \textup{\textbf{0}} \mod q$ and $\norm{\textup{\textbf{x}}} \leq \beta$. 
\end{definition}

Here, the norm $\norm{\cdot}$ is usually specified to be $\norm{\cdot}_2$ or $\inorm{\cdot}$. Observe that the restriction $\beta < q$ is necessary to avoid trivial solutions. If $\beta \geq q$, the solution $\textbf{x} = (q,0,\dots,0)\in \Z^m$ satisfies $\textbf{A}\textbf{x}\equiv \textbf{0} \mod q$ for any $\textbf{A}, m, k$. The problem in Definition \ref{def.SIS} was first introduced by Ajtai in \cite{Ajt96}, who shows the SIS problem is as at least as hard as solving some believed-to-be difficult problem for every lattice. More specifically, solving \textsf{SIS} parameterized by $m,k,q,\beta$ with $\textbf{A}$ is equivalent to solving \textsf{SVP} on the $q$-ary lattice $\mathcal{L}_q^\perp (\textbf{A})$. Furthermore, for $\textbf{A}\xleftarrow{\$} \Z_q^{m\times n}$ a solution $\textbf{x}$ with $\norm{\textbf{x}} \leq \beta$ to \textsf{SVP} in the lattice $\mathcal{L}_q^\perp (\textbf{A})$ gives a solution to $\beta\sqrt{n}$-\textsf{GapSVP} and $\beta\sqrt{n}$-\textsf{SIVP} on any $n$ dimensional lattice. We refer the reader to \cite{Ajt96} for more details.

In practical constructions of cryptosystems, a variation of SIS over modules known as module short integer solutions (\textsf{MSIS}) is used for advantages in efficiency and security. We outline the \textsf{MSIS} problem formally in Definition \ref{def:MSIS} below.

\begin{definition}[Module Short Integer Solution (MSIS)]\label{def:MSIS} 

Let $m, k, n, q$, and $\beta < q$ be positive integers. For a randomly-chosen matrix $\bA\leftarrow \mathcal{R}_q^{m\times k}$, the \textup{\textsf{MSIS}} problem is to find $\textup{\textbf{x}}\neq \textup{\textbf{0}} \in \mathcal{R}^k$ such that $\bA \textup{\textbf{x}} \equiv \textup{\textbf{0}} \mod (X^n+1,q)$ and $\norm{\textup{\textbf{x}}} \leq \beta$.
\end{definition}

We remark that although we use the specific rings $\mathcal{R}_q$ and $\mathcal{R}$, \textsf{MSIS} can be more generally defined for any polynomial $f(X)$ of degree $n-1$ for the rings $\Z_q[X]/(f(X))$ and $\Z[X]/(f(X))$ respectively. \textsf{SIS} and \textsf{MSIS} are particularly useful in cryptography to construct a one-way function (see, e.g., \cite{Ajt96,1181960,10.1561/0400000074}). In lattice-based signature schemes such as CRYSTALS-Dilithium, the the hardness of \textsf{MSIS} is needed to ensure crucial security assumptions hold such as strong unforgeability \cite{crystalsdilithium_ducas_2018}.

\subsection{Learning With Errors Problem}\label{sec:LWE}

Another popular and widely used problem in lattice-based cryptography is the Learning With Errors (\textsf{LWE}) problem. The \textsf{LWE} problem, introduced by Regev in \cite{Reg05}, tasks the solver with finding a solution to a system of noisy linear equations modulo $q$. Formally, \textsf{LWE} can be defined as follows.

\begin{definition}[Learning With Errors (LWE)]\label{def.lwe}
Let $\textup{\textbf{s}}\in \Z_q^k$ fixed and $\chi$ a probability distribution on $\Z$. Suppose $\textup{\textbf{A} }\xleftarrow{\$} \Z_q^{m\times k}$ and $\textup{\textbf{b}}\in \mathbb{Z}_q^k$ is computed as
$
\textup{\textbf{b}} = \textup{\textbf{A}} \textup{\textbf{s}} + \textup{\textbf{e}} \mod q
$
for $\textup{\textbf{e}}\leftarrow \chi^k$. 
Given $(\textup{\textbf{A}}, \textup{\textbf{b}}) \in \Z_q^{m\times k}\times \Z_q^k$, the \textup{\textsf{LWE}} problem is to recover $\textup{\textbf{s}}$.
\end{definition}

Most commonly, $\chi$ is chosen to be the discrete Gaussian distribution (i.e., $\chi = \mathcal{D}_{\sigma}$). When $\chi = \mathcal{D}_{\sigma}$ with $\sigma \geq 2\sqrt{k}$, Regev \cite{Reg05} shows solving LWE on average is at least as hard as solving some approximate lattice problems quantumly. In \cite{cryptoeprint:2008/481}, Peikert shows a similar result for a classical reduction given some restrictions on $q$. For more details on reductions relating to LWE, see e.g. \cite{Reg05,cryptoeprint:2008/481,5497885,cryptoeprint:2013/069}.

As in the case of \textsf{SIS}, \textsf{LWE} can be generalized to the module setting which we call module learning with errors (\textsf{MLWE}). We give the \textsf{MLWE} problem in Definition \ref{def:MLWE} below.

\begin{definition}[Module Learning With Errors (MLWE)]\label{def:MLWE} Let $\textup{\textbf{s}}\in \mathcal{R}_q^k$ fixed and $\chi$ a probability distribution on $\mathcal{R}$. Suppose $\textup{\textbf{A} }\xleftarrow{\$} \mathcal{R}_q^{m\times k}$ and $\textup{\textbf{b}}\in \mathcal{R}_q^k$ is computed as
$
\textup{\textbf{b}} = \textup{\textbf{A}} \textup{\textbf{s}} + \textup{\textbf{e}} \mod (X^n+1,q)
$
for $\textup{\textbf{e}}\leftarrow \chi^k$. 
Given $(\textup{\textbf{A}}, \textup{\textbf{b}}) \in \mathcal{R}_q^{m\times k}\times \mathcal{R}_q^k$, the \textup{\textsf{MLWE}} problem is to recover $\textup{\textbf{s}}$.
\end{definition}

As with \textsf{LWE}, the distribution $\chi$ in \textsf{MLWE} is commonly chosen to be discrete Gaussian (i.e., $\chi = \mathcal{D}_{\sigma}^n$). In a similar to fashion to \textsf{MSIS}, \textsf{MLWE} is a well-known and extensively used problem in several areas of lattice-based cryptography. Most relevant to our work in this paper, the hardness of \textsf{MLWE} forms a significant foundation for security against key recovery attacks in signature schemes, such as CRYSTALS-Dilithium \cite{crystalsdilithium_ducas_2018}.

We should note that in Definitions \ref{def.lwe} and \ref{def:MLWE}, we respectively phrase \textsf{LWE} and \textsf{MLWE} as the problem of recovering a secret vector \textbf{s}. This is known as the \textit{search} version of \textsf{LWE} or \textsf{MLWE}. However, the two problems could instead be given as a \textit{decision} problem in which one must decide if $(\textup{\textbf{A}}, \textup{\textbf{b}})$ is constructed as described or sampled uniform randomly. Though we do not outline the decision version of the problems in this paper, solving the search version and the decision version of \textsf{LWE} (or \textsf{MLWE}) is actually equivalent.

A lesser known variant of the \textsf{LWE} problem which omits the modular reduction is introduced by Bootle et al. in \cite{bootle}. This variant, known as integer learning with errors (\textsf{ILWE}), is given below as Definition \ref{def:ILWE}.

\begin{definition}[Integer Learning With Errors (ILWE)]  \label{def:ILWE} 
Let $\textup{\textbf{s}}\in \Z^k$ fixed and $\chi_a,\chi_e$ probability distributions on $\Z$. Suppose $\textup{\textbf{A} }\leftarrow \chi_a^{m\times k}$ and $\textup{\textbf{b}}\in \mathbb{Z}^k$ is computed as
$
\textup{\textbf{b}} = \textup{\textbf{A}} \textup{\textbf{s}} + \textup{\textbf{e}}
$
for $\textup{\textbf{e}}\leftarrow \chi_e^k$. 
Given $(\textup{\textbf{A}}, \textup{\textbf{b}}) \in \Z^{m\times k}\times \Z^k$, the \textup{\textsf{ILWE}} problem is to recover $\textup{\textbf{s}}$.
\end{definition}

In comparison to the construction in \textsf{LWE}, observe that the entries of \textbf{b} = \textbf{A}\textbf{s} + \textbf{e} are not reduced modulo $q$ in \textsf{ILWE}. The reduction modulo $q$ is a crucial step for ensuring the hardness of \textsf{LWE}, hence the \textsf{ILWE} problem is less used in the foundations of forming practical lattice-based cryptographic protocols. However, when paired with rejection sampling (to be discussed in Section \ref{section.rejection_sampling}), \textsf{ILWE}-like constructions appear in lattice-based digital signatures such as BLISS and Dilithium.

\subsection{Attacks on ILWE via Least Squares}\label{section.LS_attack}

After introducing the ILWE problem, \cite{bootle} develops and outlines an effective attack against \textsf{ILWE} based on the linear least squares method (LSM). The attack is simple and elegant, yet quite effective so long as the number of rows $m$ in the \textsf{ILWE} instance is large enough and the error distribution $\chi_e$ is \textit{subgaussian}:

\begin{definition}[Subgaussian Variable]
Let $\tau>0$ be a real number. A random variable $X$ over $\mathbb{R}$ is said to be $\tau$-subgaussian if for all $u \in \mathbb{R}$,
$$\mathbb{E} [\exp(uX)]\leq \exp \left(\frac{\tau^2u^2}{2}\right).$$ 
A $\tau$-subgaussian probability distribution is defined similarly.
\end{definition}

The term subgaussian is a broad term that applies to several probability distributions common in \textsf{LWE} and \textsf{ILWE}. For instance, the discrete Gaussian distribution $\mathcal{D}_{\sigma}$ is $\frac{\sigma}{\sqrt{2\pi}}$-subgaussian 
and the uniform distribution over $[-\alpha,\alpha]\cap \Z$ is $\frac{\alpha}{\sqrt{2}}$-subgaussian  
\cite{bootle}. We now summarize the LSM attack below as Algorithm \ref{alg.ls}, which we will commonly refer to as the ``LSM attack".

\begin{figure*}[h!]
   \begin{center}
   \renewcommand{\arraystretch}{1.25}
       \begin{tabular}{|p{1.3cm}p{10.5cm}p{.02cm}|}
       \hline
       & $\tx{Attack.LSM}(\textbf{A},\textbf{b})$ &\\
       \hline
       Input: & $\textbf{A}\in \mathbb{R}^{m\times n}$ and $\textbf{b}\in \mathbb{R}^{m}$, matrices for an ILWE instance. &\\
       Output: & $\tilde{\textbf{s}}\in \mathbb{R}^{n}$ recovered secret. &\\
        \hline
            Step 1. & Compute $\hat{\textbf{s}} := (\textbf{A}^\top \textbf{A})^{-1} \textbf{A}^{\top} \textbf{b}$. &\\
            Step 2. & Compute $\tilde{\textbf{s}} = \lfloor \hat{\textbf{s}}\rceil$. &\\
            Step 3. & Return $\tilde{\textbf{s}}$. &\\
      \hline
      \end{tabular}
      \renewcommand{\arraystretch}{1}
      \caption{LSM Attack on \textsf{ILWE}}
      \label{alg.ls}
   \end{center}
\end{figure*}

The authors in \cite{bootle} show that if given sufficiently many \textsf{ILWE} samples (i.e., sufficiently many rows $m$ of $\textbf{A})$, then the attack in Algorithm \ref{alg.ls} obtains $\hat{\textbf{s}}$ satisfying $\inorm{\hat{\textbf{s}} - \textbf{s}} < 1/2$ with high probability, hence $\tilde{\textbf{s}} = \textbf{s}$. For the distributions $\chi_a$ and $\chi_e$, let us denote $\sigma_a$ and $\sigma_e$ as their standard deviations respectively. We then have the  Theorem \ref{thm.bootle} below adapted from Theorem 4.5 in \cite{bootle}.

\begin{theorem}[adapted from \cite{bootle}]\label{thm.bootle}

Suppose that $\chi_a$ is $\tau_a$-subgaussian and $\chi_e$ is $\tau_e$-subgaussian, and let $(\textup{\textbf{A}},\textup{\textbf{b}} = \textup{\textbf{A}}\textup{\textbf{s}} + \textup{\textbf{e}})$ the data constructed from $m$ samples of the ILWE with $\textup{\textbf{A}}\leftarrow \chi_a^{m\times k}$, $\textup{\textbf{e}} \leftarrow \chi_e^{k}$, and $\textup{\textbf{s}}\in \Z^k$. Then, there exists constants $C_1,C_2 > 0$ such that for all $\eta \geq 1$, if:
\[
m \geq 4 \frac{\tau_a^4}{\sigma_a^4}(C_1 k + C_2 \eta) \quad \text{and} \quad m\geq 32 \frac{\tau_e^2}{\sigma_a^2}\log (2k)
\]
then the least squares estimator $\hat{\textup{\textbf{s}}} = (\textup{\textbf{A}}^{\top} \textup{\textbf{A}})^{-1} \textup{\textbf{A}}^\top \textup{\textbf{b}}$ satisfies $\inorm{\textup{\textbf{s}} - \hat{\textup{\textbf{s}}}} < 1/2$, and hence $\tilde{\textup{\textbf{s}}} = \textup{\textbf{s}}$, with probability at least $1 - \frac{1}{2n} - 2^{-\eta}$.
\end{theorem}

It can be shown that $C_1= 2^8 \log 9$ and $C_2 = 2^9 \log 2$ in Theorem \ref{thm.bootle}. The result of this theorem indicates that the attack in Algorithm \ref{alg.ls} actually does very well against \textsf{ILWE}. In fact, the authors of \cite{bootle} use this attack in conjunction with information obtained via side-channel attacks to fully recover secret keys in the BLISS signature scheme (to be discussed in Section \ref{section.BLISS}). Although we take a different approach in our analyses, this highlights the effectiveness of a reasonable simple attack strategy against \textsf{ILWE}.

We should emphasize that this attack exploits the underlying mathematical structure of information known (or deduced) by a party, and hence we consider it as a digital attack. Although we will discuss the implications of this work and the broader impact in practical applications in Section \ref{section.applications}, no consideration to physical components will be considered. Our later experiments conducted with this attack, outlined in Section \ref{section.concretesecurity}, use only publicly available digital information.

\subsection{Attacks on ILWE via Singular Value Decomposition}\label{section.svd}

An additional attack we describe in this paper comes from Gao \cite{gao2025boundeddistancedecodingrandom}, which uses a singular value decomposition (SVD) to solve bounded distance decoding. This algorithm may also be recover the secret $\textbf{s}$ in an \textsf{ILWE} instance. Although not the central focus of our work, we opt to describe the basics of this attack since it has not been previously studied in this context. We provide a simplified version of the algorithm in Algorithm \ref{alg.svd} below, and refer the reader to \cite{gao2025boundeddistancedecodingrandom} for a more comprehensive view of the algorithm.

\begin{figure*}[h!]
   \begin{center}
   \renewcommand{\arraystretch}{1.25}
       \begin{tabular}{|p{1.3cm}p{10.5cm}p{.02cm}|}
       \hline
       & $\tx{Attack.SVD}(\textbf{A},\textbf{b})$ &\\
       \hline
       Input: & $\textbf{A}\in \mathbb{R}^{m\times n}$ and $\textbf{b}\in \mathbb{R}^{m}$, matrices for an ILWE instance. &\\
       Output: & $\tilde{\textbf{s}}\in \mathbb{R}^{n}$ recovered secret. &\\
        \hline
            Step 1. & Form the matrix $\textbf{M}:= (\textbf{A}, -\textbf{b})$ and compute and SVD $\textbf{M} = \textbf{U} \Sigma \textbf{V}^\top$. &\\
            Step 2. & Let $\textbf{v} = (v_1,\dots,v_{n+1})^\top$ be the last column of \textbf{V}. If $v_{n+1} = 0$, abort. Else, continue. &\\
            Step 3. & For $i=1$ to $n$ compute &\\
            & \quad $\tilde{s}_i = \lfloor \frac{v_i}{v_{n+1}} \rceil \in \Z$. &\\
            Step 4. & Return $\tilde{\textbf{s}} = (\tilde{s}_1,\dots,\tilde{s}_{n})^\top$. &\\
      \hline
      \end{tabular}
      \renewcommand{\arraystretch}{1}
      \caption{SVD Attack on \textsf{ILWE}}
      \label{alg.svd}
   \end{center}
\end{figure*}

\section{Rejection Sampling and Digital Signatures}\label{section.rejection_sampling}

Rejection sampling, introduced in \cite{vonNeumann1951}, is a common technique used in several digital signature algorithms to ensure independence of the secret key from signature components. The basic idea is to sample elements from some probability distribution and reject them with some probability. More specifically, for a given probability distribution $f$ and for some positive constant $M\in \mathbb{R}$, a sample $u$ is drawn from $f$ and accepted with probability $g(u)/(M \cdot f(u))$ for some other probability distribution $g$. This ensures that our samples actually follow the distribution $g$. This allows one to generate samples following the distribution $g$, while only accessing samples from the distribution $f$. Although this describes the basic idea of rejection sampling, we refer the reader to other works \cite{vonNeumann1951,cryptoeprint:2013/383} for a more detailed and comprehensive foundation for rejection sampling.

\subsection{CRYSTALS-Dilithium}

One of the most popular post-quantum digital signature schemes is CRYSTALS-Dilithium \cite{crystalsdilithium_ducas_2018}. In fact, the module-lattice-based digital signature standard (ML-DSA) standardized by NIST in 2022 \cite{NIST:FIPS204} is derived from CRYSTALS-Dilithium. The overall security of CRYSTALS-Dilithium is based on two well-known lattice problems: the MSIS \eqref{def:MSIS} and MLWE \eqref{def:MLWE} problems. 
In the main signing algorithm, a rejection sampling procedure is used. We provide an overview of said rejection sampling loop below in Algorithm \ref{Dilithium.rejection.sampling}. Note that we omit a large majority of the actual signing procedure for simplicity, as we will only be concerned with this rejection sampling procedure. The full signing algorithm can be found in \cite{crystalsdilithium_ducas_2018,NIST:FIPS204}.

\begin{figure*}[h!]
   \begin{center}
   \renewcommand{\arraystretch}{1.25}
       \begin{tabular}{|p{1.3cm}p{12.5cm}p{.02cm}|}
       \hline
       & $\tx{Dilithium.Samples}(k,\textbf{s},\gamma,\rho,\beta)$ &\\
       \hline
       Input: & $k \in \mathbb{N}$ module rank, &\\ 
       & $\textbf{s} \in \mathcal{R}^k$ secret with coefficients in $[-\eta,\eta]$, &\\
       & $\gamma \in \mathbb{N}$ parameter satisfying $\inorm{\textbf{y}} \leq \gamma$, &\\ & $\rho \in \mathbb{N}$ Hamming weight of $c$, & \\ & $\beta \in \mathbb{N}$ parameter for rejection sampling bound. &\\
        \hline
            Step 1. & Sample uniform random $\textbf{y} \leftarrow \mathcal{R}_q^k$ with each polynomial having coefficients in $[-\gamma+1,\gamma]$. &\\
            Step 2. & Sample random $c\leftarrow \mathcal{R}_q$ with coefficients in $\{0,\pm 1 \}$ and Hamming weight $\rho$. &\\
            Step 3. & Compute $\textbf{z} : = \textbf{y} + c\textbf{s} \in \mathcal{R}^{k}$. &\\
            Step 4. & Reject if $\norm{\textbf{z}}_{\infty} \geq \gamma - \beta$ and go back to Step 2. Else, continue. &\\
            Step 5. & Return $(\textbf{z},c)$.&\\
      \hline
      \end{tabular}
      \renewcommand{\arraystretch}{1}
      \caption{Basic Overview of Dilithium's Rejection Sampling Procedure}
      \label{Dilithium.rejection.sampling}
   \end{center}
\end{figure*}

We provide some explanations and important observations for a few of these steps. In Step 2, we need to randomly sample an element of $\mathcal{R}_q$ with coefficients in $\{0,\pm 1 \}$ and Hamming weight $\rho$. The standard method for this is to use the \tx{SampleInBall} procedure outlined below in Algorithm \ref{alg.sampleinball} as done in \cite{crystalsdilithium_ducas_2018}. This procedure (or similarly designed procedures) are fairly standard for hashing to a ball. In our later experiments, we use a basic implementation of this procedure to generate $c\in \mathcal{R}$ by simply generating the coefficients vector of $c$ via Algorithm \ref{alg.sampleinball}. Careful consideration and precaution should be used when implementing or using this procedure in practice however.

\begin{figure*}[h!]
   \begin{center}
   \renewcommand{\arraystretch}{1.25}
       \begin{tabular}{|p{1.3cm}p{8.5cm}p{.02cm}|}
       \hline
       & $\tx{SampleInBall}(\rho)$ &\\
       \hline
       Input: & $\rho \in \mathbb{N}$ parameter. &\\
       Output: & $\textbf{c}\in \{0,\pm 1 \}^{n}$ with Hamming weight $\rho$. &\\
        \hline
            Step 1. & Initialize vector $\textbf{c}=\textbf{0}$ of length $n$. & \\
            Step 2. & For $i:=(n+1)-\rho$ to $n$ do &\\
            & \quad $j\xleftarrow{\$}\{1,\dots,i\}$, &\\
            & \quad $\textbf{c}[i] = \textbf{c}[j]$, &\\
            & \quad $\alpha \xleftarrow{\$} \{0,1\}$, &\\
            & \quad $\textbf{c}[j] = (-1)^{\alpha}$. &\\
            Step 3. & Return $\textbf{c}$. &\\
      \hline
      \end{tabular}
      \renewcommand{\arraystretch}{1}
      \caption{Creating a Random Vector with $\rho$ Entries $\pm 1$.}
      \label{alg.sampleinball}
   \end{center}
\end{figure*}

For Step 3, note that we compute $\textbf{z}$ as an element in $\mathcal{R}^k$ (i.e., without reducing coefficients modulo $q$). Although many specifications note that $\textbf{z}$ should be computed modulo $q$, it actually does not make a difference. Since $\inorm{\textbf{y}} \leq \gamma$ and $\inorm{c \textbf{s}}\leq \rho\cdot \eta$, it is clear that $\inorm{\textbf{y} + c\textbf{s}} \leq \gamma + \rho\cdot \eta < q/2$ for essentially any reasonable choice of parameters. Hence reduction modulo $q$ does not occur even when computing in $\mathcal{R}_q^k$. 

In Step 4, the value $\gamma-\beta$ serves as the rejection sampling bound. We should note that the parameter $\beta$ is usually chosen as the maximum possible coefficient of $c\textbf{s}$ for verification to be correct. In our later experiments, we will consider a slightly different case since we only are concerned with security of this key recovery attack. Instead of focusing solely on $\beta$, we will consider the more general bound $\gamma-\beta$ which appears in the rejection sampling. For fixed $\gamma$, we choose $\beta$ so that the rejection rate is roughly $50\%$ (more specifically, between $40\%$ and $60\%$ most of the time). The change in $\beta$ is most noticeable for smaller parameter experiments which is discussed in Section \ref{section.concretesecurity}, where the value of $\beta$ may sometimes even be negative. This is acceptable for use in our simulations. We also slightly adjust $\rho$ in our experiments, the Hamming weight of $c$ for different scenarios.

\subsection{BLISS}\label{section.BLISS}

In \cite{bootle}, which introduces the attack described in Section \ref{section.LS_attack}, the authors apply the LSM attack to BLISS (Bimodal Lattice Signature Scheme) \cite{cryptoeprint:2013/383}. Though the structure of BLISS has a quite similar structure to Dilithium, \cite{bootle} applies the LSM attack in a very different way than we will. Our formulation of an \textsf{ILWE} instance will be discussed later in Section \ref{subsection.arithmetic}. \cite{bootle} apply the attack to BLISS by using a known value of $\< \textbf{z}, c\textbf{s}\>$ obtained via side-channel attacks. To best describe their approach, we introduce the main signing algorithm from BLISS (Algorithm \ref{algorithm.bliss}) which the attack is used on.

\begin{figure}[h!]
   \begin{center}
   \renewcommand{\arraystretch}{1.25}
    \renewcommand{\figurename}{Algorithm}
       \begin{tabular}{|p{1.3cm}p{11.5cm}p{.02cm}|}
       \hline
       & $\texttt{BLISS.Sign}(\textbf{s},\tx{pk},m,\zeta)$ &\\
       \hline
       Input: & $\textbf{s} = (s_1,s_2) \in \mathcal{R}_{2q}^2$ secret, &\\
       & $\tx{pk}\in \mathcal{R}_{2q}$ public key, &\\
       & $m \in \mathcal{R}_{q}$ message,&\\
       & $\zeta \in [0,2q-1]$ with $\zeta\cdot (q-2)=1  \mod 2q$. &\\
       
       Output: & $(\textbf{z} = (z_1,z_2'),c)\in \mathcal{R}_{2q}^2\times \mathcal{R}$ signature. &\\
        \hline
            Step 1. & Sample $y_1,y_2 \leftarrow \mathcal{D}_{\sigma}^{n}$. &\\
            Step 2. & Compute $u:= \zeta \cdot \tx{pk}\cdot y_1 + y_2 \mod 2q$.   &\\
            Step 3. & Compute $c := H(\lfloor u \rceil_d \mod p, m) \in \mathcal{R}$. &\\
            Step 4. & Sample $b\xleftarrow{\$} \{ 0,1\}$ and compute  &\\
            & \quad $z_1:= y_1 + (-1)^b s_1 c$, &\\
            & \quad $z_2:= y_2 + (-1)^b s_2 c$. &\\
            Step 5. & Continue with probability $1/(M \text{exp} (-\norm{\textbf{s}c}_2^2 / (2\sigma^2)) \text{cosh}(\<\textbf{z},\textbf{s}c \>) )$. Else, go back to Step 1. &\\
            Step 6. & Compute  $z_2':= (\lfloor u\rceil_d - \lfloor u - z_2 \rceil_d) \mod p$. &\\
            Step 7. & Return $(\textbf{z} = (z_1,z_2'),c)$. &\\
      \hline
      \end{tabular}
      \renewcommand{\arraystretch}{1}
      \caption{BLISS Signing Algorithm}
      \label{algorithm.bliss}
   \end{center}
\end{figure}

We omit several finer details of Algorithm \ref{algorithm.bliss}. This includes generation of $\tx{pk}$, the notation $\lfloor \cdot \rceil_d$, and some additional parameters like $M$ and $p$. $H$ is a hash to a ball (e.g., \tx{SampleInBall}). We refer the reader to \cite{cryptoeprint:2013/383} for a more comprehensive outlook on the BLISS signing algorithm. Observe however that the internal structure of the signing algorithm is quite similar to Dilithium, computing 
\begin{align*}
   z_1&:= y_1 + (-1)^b s_1 c,\\ 
   z_2&:= y_2 + (-1)^b s_2 c,
\end{align*}
with $\textbf{z} = (z_1,z_2)$, $\textbf{y} = (y_1,y_2)$, and $\textbf{s} = (s_1,s_2)$. That is, $\textbf{z}\in \mathcal{R}^2$ satisfies
\[
\textbf{z} := \textbf{y} \pm c \textbf{s} \in \mathcal{R}^2. 
\]
Though there are several other differences between the signing algorithms, we take particular notice of the rejection sampling step. In BLISS, signatures are rejection with some probability, whereas Dilithium rejects them based on the infinity norm of part of the sample.

To formulate an \textsf{ILWE} instance based on the structure of BLISS signatures, \cite{bootle} show that $\< \textbf{z}, c\textbf{s}\> = \< \textbf{a}, \textbf{s}\> + e$ with
\[
\textbf{a} = (z_1 c^*, 2^d z_2' c^*) \quad \text{and}\quad e = \< z_2 - 2^d z_2', s_2 c\>.
\]
Here, $c^*$ is the conjugate of $c$ with respect to the inner product. For many samples, an \textsf{ILWE} may be constructed with the matrix \textbf{A} with rows consisting of each $\textbf{a}$ and the entries of $\textbf{b}$ consisting of each $\< \textbf{z}, c\textbf{s}\>$ recovered from side-channel attacks.

It can be shown that the $e$ described above follows some bounded distribution $\chi_e$ with variance
\[
\sigma_e^2 \approx \frac{2^{2d}}{3}(\delta_1 + 4\delta_2)n\rho
\]
for some $\delta_1$ and $\delta_2$. On the other hand, \textbf{a} follows some distribution $\chi_{a}$ with variance
\[
\sigma_a^2 = \rho \cdot \text{Var}(\mathcal{D}_{\sigma})
\]
for the underlying discrete gaussian distribution $\mathcal{D}_{\sigma}$. The main observation to make is that generally we will have $\sigma_e^2 > \sigma_a^2$. This is true for all the example parameter choices used in \cite{bootle}.

\section{Implementation Strategies}\label{section.implementation_strategies}

In this section, we outline our techniques for implementing our attack simulations. We specifically apply the attacks to outputs of Algorithm \ref{Dilithium.rejection.sampling}. Said otherwise, given many samples of the form $(\textbf{z}_i,c_i)\in \mathcal{R}^k$ with
\begin{equation}\label{eq.main}
\textbf{z}_i = \textbf{y}_i + c_i
\textbf{s}\in \mathcal{R}^k
\end{equation}
we wish to attempt to recover \textbf{s} using the attack from Bootle et al
and Gao. The attacks have been implemented in release R2023b (version 23.2) of \tx{MATLAB}, which utilizes matrix operations over $\mathbb{R}$. For this to be successfully done, there are two techniques we use: converting polynomial arithmetic in $\mathcal{R}$ to matrices over $\mathbb{R}$, and an algorithms for handling large sample sizes.

\subsection{Arithmetic in $\mathcal{R}$ via Matrices Over $\R$}\label{subsection.arithmetic}

In formulating the \textsf{ILWE} instance and discussing algorithms of implementing the described attack, it will be useful to consider the problem of recovering \textbf{s} with many samples in the form of equation \eqref{eq.main} as a problem over integer (and real) matrices. For any two polynomials $f\in \mathcal{R}$ and $g\in \mathcal{R}$, write
\begin{align*}
    f &= f_0 + f_1 X + \dots + f_{n-1}X^{n-1},\\
    g &= g_0 + g_1 X + \dots + g_{n-1}X^{n-1}.
\end{align*}
Define the matrix $\textbf{F}\in \Z^{n\times n}$ and vector $\textbf{G} \in \Z^{n}$ via
\[
\textbf{F} = \begin{bmatrix}
    f_0 & -f_{n-1} & -f_{n-2} & -f_{n-3} & \cdots & -f_1 \\
    f_1 & f_{0} & -f_{n-1} & -f_{n-2} & \cdots & -f_2 \\
    f_2 & f_{1} & f_{0} & -f_{n-1} & \cdots & -f_3 \\
    \vdots & \vdots & \vdots & \vdots & \ddots & \vdots \\
    f_{n-1} & f_{n-2} & f_{n-3} & f_{n-4} &\cdots & f_{0}
\end{bmatrix}, \quad \textbf{G} = \begin{bmatrix}
    g_0 \\
    g_1 \\
    g_2 \\
    \vdots\\
    g_{n-1}
\end{bmatrix}.
\]
Here, $\textbf{F}$ is simply the  matrix representation of $f\in \mathcal{R}$ under the ordered monomial basis $(1,X,\dots,X^{n-1})$ of $\mathcal{R}$. We then claim that we can compute the coefficients of $fg \mod X^n+1$ by the matrix-vector product using the defined \textbf{F} and \textbf{G}. That is,
\[
\tx{coeff}(fg \mod X^n+1)  = \textbf{F}\textbf{G}.
\]
In practice, computing the matrix-vector product is actually less efficient than computing the polynomial product and reducing modulo $X^n+1$. Computing $fg \mod X^n+1$ takes $O(n\log n)$ operations using a number theoretic transform (see, e.g., \cite{cryptoeprint:2024/585}) whereas the product $\textbf{F}\textbf{G}$ is naively $O(n^2)$. This matrix-vector product will be used for crafting our \textsf{ILWE} instance however.

We can extend the above ideas to the product $f\textbf{g}\in \mathcal{R}^k$ for $f\in \mathcal{R}$ and $\textbf{g}\in \mathcal{R}^k$ in the following way. Write
\[
\textbf{g} = (g^{(0)},g^{(1)},\dots,g^{(k-1)}) \in \mathcal{R}^k
\]
and let $\textbf{G}^{(j)}$ be the coefficient vector of $g^{(j)}$. Define matrix $\tilde{\textbf{F}}\in \Z^{nk\times nk}$ and vector $\tilde{\textbf{G}} \in \Z^{nk}$ via
\[
\tilde{\textbf{F}} = \begin{bmatrix}
    \textbf{F} & & & \\ 
     & \textbf{F} & & \\
     & & \ddots & \\
     & & & \textbf{F}
\end{bmatrix}, \quad \tilde{\textbf{G}} = \begin{bmatrix}
    \textbf{G}^{(0)} \\
    \textbf{G}^{(1)} \\
    \vdots \\
    \textbf{G}^{(n-1)}
\end{bmatrix}.
\]
Then,
\[
\tx{coeff}(f\textbf{g} \mod X^n+1) = \tilde{\textbf{F}}\tilde{\textbf{G}}.
\]
In regards to our specific problem, observe that each sample $(\textbf{z}_i,c_i)\in \mathcal{R}_{q}^k\times \mathcal{R}_{q}$ may be represented by a matrix $\tilde{\textbf{C}}_i\in \Z^{nk\times nk}$ and a vector $\tilde{\textbf{Z}}_i\in \Z^{nk}$. Supposing we have access to $m$ samples, recovering $\textbf{s}$ can then be done by solving an \textsf{ILWE} instance with
\[
\textbf{A} =
\begin{bmatrix}
    \tilde{\textbf{C}}_1\\
    \tilde{\textbf{C}}_2\\
    \vdots\\
    \tilde{\textbf{C}}_m
\end{bmatrix} \in \Z^{mnk\times nk}, \quad \textbf{b} = \begin{bmatrix}
    \tilde{\textbf{Z}}_1 \\
    \tilde{\textbf{Z}}_2 \\
    \vdots \\
    \tilde{\textbf{Z}}_m
\end{bmatrix} \in \Z^{mnk}.
\]

\subsection{Handling Large Sample Sizes}\label{section.largesamples}

The matrix \textbf{A} for the \textsf{ILWE} instance described has $mnk$ rows, which clearly will be difficult to deal with for large sample sizes $m$. Computing the least squares estimator for a matrix of size $mnk\times nk$ can be done in  $O(mn^2 k^2)$. Below we show an algorithm with this complexity (Algorithm \ref{largesamples}), specifically catered to the structure of samples in $\mathcal{R}$ with real matrices.

\begin{figure}[h!]
   \begin{center}
   \renewcommand{\arraystretch}{1.25}
    \renewcommand{\figurename}{Algorithm}
       \begin{tabular}{|p{1.3cm}p{10.5cm}p{.02cm}|}
       \hline
       & $\texttt{Samples.LSM}(m)$ &\\
       \hline
       Input: & $m$ number of samples used. &\\
       Output: & $\textbf{x}$ least squares estimator obtained for $m$ samples. &\\
        \hline
            Step 1. & Initialize the $(kn+1)\times (kn+1)$ matrix $\textbf{B}:=\textbf{0}$. &\\
            Step 2. & For $i:=1$ to $m$ do &\\
            & \quad $(a)$ Get Sample $\textbf{D}: = (\tilde{\textbf{C}} || \tilde{\textbf{Z}}) \in \R^{kn\times (kn+1)}$, & \\
            & \quad $(b)$ Compute $\textbf{B}:= \textbf{B} + \textbf{D}^\top \textbf{D}$. & \\
            Step 3. & Solve $\textbf{B}[1\text{:}kn,1\text{:}kn] \textbf{x} = \textbf{B}[1\text{:}kn,kn+1]$. &\\
            Step 4. & Return $\textbf{x}$. &\\
      \hline
      \end{tabular}
      \renewcommand{\arraystretch}{1}
      \caption{LSM Algorithm for Large Samples}
      \label{largesamples}
   \end{center}
\end{figure}
In Step 2 of Algorithm \ref{largesamples}, $(\tilde{\textbf{C}} || \tilde{\textbf{Z}})$ is the matrix representation of the sample $(\textbf{z},c)$. Recall also that $\textbf{B}[1\text{:}kn,1\text{:}kn]$ denotes the submatrix of $\textbf{B}$ consisting of the first $kn$ rows and first $kn$ columns and $\textbf{B}[1\text{:}kn,kn+1]$ the submatrix consisting of the first $kn$ rows and the last column of \textbf{B}. We now give a brief description of why Algorithm \ref{largesamples} works. For an \textsf{ILWE} instance with $\textbf{A}$ and $\textbf{b}$, let $\textbf{M} = (\textbf{A} || \textbf{b})$. Then note that
\[
\textbf{M}^\top \textbf{M} = \begin{bmatrix}
    \textbf{A}^\top \textbf{A} & \textbf{A}^\top \textbf{b} \\
    \textbf{b}^\top \textbf{A} & \textbf{b}^\top \textbf{b}
\end{bmatrix}.
\]
If 
\[
\textbf{M} = \begin{pmatrix}
    \textbf{B}_1\\
    \textbf{B}_2\\
    \vdots\\
    \textbf{B}_m\\
\end{pmatrix},
\] 
then we have $\textbf{M}^\top \textbf{M} = \sum_{i=1}^m \textbf{B}_i^\top \textbf{B}_i$. We can compute $\textbf{x} = (\textbf{A}^\top \textbf{A})^{-1} \textbf{A}^{\top} \textbf{b}$ by solving a linear system with portions of $\textbf{M}^\top \textbf{M}$. This gives the Algorithm \ref{largesamples} above, as each $\textbf{B}_i$ here is constructed with the matrix representations of the samples $\textbf{B}_i = (\tilde{\textbf{C}}_i || \tilde{\textbf{Z}}_i)$.

\subsubsection{Large Sampling Algorithms for SVD} Algorithm \ref{largesamples} can be easily modified to work for the attack introduced by Gao described in Section \ref{section.svd} as well. We give the modification as a separate algorithm in Algorithm \ref{largesamples2}.

\begin{figure}[h!]
   \begin{center}
   \renewcommand{\arraystretch}{1.25}
    \renewcommand{\figurename}{Algorithm}
       \begin{tabular}{|p{1.3cm}p{10.5cm}p{.02cm}|}
       \hline
       &$ \texttt{Samples.SVD}(m)$ &\\
              \hline
       Input: & $m$ number of samples used. &\\
       Output: & $\textbf{v}$ the last column of \textbf{V} in SVD obtained for $m$ samples. &\\
        \hline
            Step 1. & Initialize the $(kn+1)\times (kn+1)$ matrix $\textbf{B}:=\textbf{0}$. &\\
            Step 2. & For $i:=1$ to $m$ do &\\
            & \quad $(a)$ Get Sample $\textbf{D}: = (\tilde{\textbf{C}} || -\tilde{\textbf{Z}}) \in \R^{kn\times (kn+1)}$, & \\
            & \quad $(b)$ Compute $\textbf{B}:= \textbf{B} + \textbf{D}^\top \textbf{D}$. & \\
            Step 3. & Compute the SVD $\textbf{B} = \textbf{V} \Sigma \textbf{V}^{\top}$. &\\
            Step 4. & Return $\textbf{v} = \textbf{V}[1\text{:}kn+1,kn+1]$, the last column of $\textbf{V}$. &\\
      \hline
      \end{tabular}
      \renewcommand{\arraystretch}{1}
      \caption{SVD Algorithm for Large Samples}
      \label{largesamples2}
   \end{center}
\end{figure}
For an \textsf{ILWE} instance, suppose the matrix $\textbf{M} = (\textbf{A}|| -\textbf{b})$ has an SVD $\textbf{M} = \textbf{U}\Sigma \textbf{V}^T$. Then an SVD of $\textbf{M}^\top \textbf{M}$ is given by
\[ \textbf{M}^\top \textbf{M} = \textbf{V}\Sigma^\top \Sigma \textbf{V}^\top.
\]
In fact, any SVD of $\textbf{M}^\top \textbf{M}$ gives rise to a matrix $\textbf{V}$ for an SVD of $\textbf{M}$. So, we only need to compute $\textbf{M}^\top \textbf{M}$, then compute an SVD of $\textbf{M}^\top \textbf{M}$ to run the attack. Though we do not outline the experimental results of this algorithm in our paper, we note that Algorithm \ref{largesamples2} can in fact be used to apply the attack from Algorithm \ref{alg.svd} for large sample instances of \textsf{ILWE}.

\section{Experimental Model and Results}\label{section.concretesecurity}

\subsection{Attack Model}

The strategies outlined in Sections \ref{subsection.arithmetic} and \ref{section.largesamples} are used to simulate the two attacks outlined in Section \ref{section.LS_attack}. We begin by describing our overall attack model used to evaluate the effectiveness of these attacks. Our workflow for these simulations follows the general framework below:
\begin{enumerate}
    \item Fix parameters $n,\rho,k,\eta$, randomly choose a secret $\textbf{s}\in \mathcal{R}^k$ with coefficients in $[-\eta,\eta]$, and set $\gamma,\beta$ so that the rejection sampling bound we reject roughly about 50\% of our samples (through trial and error).

    \item For $m=10^2$, $10^3$, $10^4$, and $10^5$, obtain $m$ samples $(\textbf{z},c)$ in matrix representation  using the implementation techniques from Section \ref{section.implementation_strategies}.

    \item Run the attack from Section \ref{section.LS_attack} to obtain our recovered $\tilde{\textbf{s}}$.

    \item Compute and return $\norm{\tilde{\textbf{s}} - \textbf{s}}_1$, discarding the result if $\tilde{\textbf{s}} = \textbf{0}$.
\end{enumerate}

The result we obtain with the above workflow is $\norm{\tilde{\textbf{s}} - \textbf{s}}_1$. We choose the 1-norm to measure the ``success" of an attack in an attempt to balance the number of slots recovered with the sizes of coefficients recovered. Notice furthermore that $\tx{wt}(\tilde{\textbf{s}} - \textbf{s}) \leq \norm{\tilde{\textbf{s}} - \textbf{s}}_1$ and $\norm{\tilde{\textbf{s}} - \textbf{s}}_{\infty} \leq \norm{\tilde{\textbf{s}} - \textbf{s}}_1$, so this result actually serves as a worst-case bound for both the difference in the largest coefficient and the number of slots recovered. If $\rho$ and $\eta$ are known, an obtained result $\tilde{\textbf{s}}$ can be adjusted so that $\tilde{\textbf{s}}$ has the correct weighting and its coefficients are in the correct range.

As mentioned in Section \ref{section.implementation_strategies}, the attacks have been implemented in release R2023b (version 23.2) of \tx{MATLAB}. We use a 13th Gen Intel Core i7-1355U 1.70 GHz processor to run the simulations. We should note however that the processor is largely irrelevant for these experimental results. Since we obtain a specific $\tilde{\textbf{s}}$ resulting from the attack which we define out success from, processing power does not matter. A better processor will allow us to run the attack with more samples (which may result in better $\tilde{\textbf{s}}$), however the resulting $\tilde{\textbf{s}}$ will be roughly the same so long as the same number of samples are used.

\subsection{Results}

We outline our simulation results in Tables \ref{tab:securityparameters1} to \ref{tab:securityparameters3} below. Several trials have been conducted for each parameter combination and each choice of $m$, with the best result (i.e., smallest $\norm{\tilde{\textbf{s}} - \textbf{s}}_1$) shown in the table. The best result is usually for $m=10^5$ samples. In Table \ref{tab:securityparameters1}, \textbf{y} is generated according to a more general subgaussian distribution. Specifically, for a parameter $\alpha$, \textbf{y} is generated in the following way. First, $\textbf{v} = (v^{(0)},\dots,v^{(k-1)})\leftarrow \mathcal{R}^k$ randomly, with each $v^{(j)}$ having Hamming weight $\rho$ and coefficients in $[-\alpha,\alpha]$. $\textbf{y}$ is then computed as
\[
\textbf{y} = \sum_{j=1}^{\rho} b_j (X^{r_j} \textbf{v} \mod X^n+1),
\]
where each $b_j \xleftarrow{\$} \{ -1,1\}$ and each $r_j \xleftarrow{\$} \{0,1,\dots,n-1\}$. 
When computed in this way, the entries of  $\textbf{y}$ follow a $(\alpha \sqrt{\rho})$-subgaussian distribution.
In Tables \ref{tab:securityparameters2} and \ref{tab:securityparameters3}, the sampling procedure for \textbf{y} is simpler. \textbf{y} is sampled so that all coefficients are uniform random from $[-\gamma,\gamma]\cap \Z$. When sampled in this way, the entries of $\textbf{y}$ follow a $(\gamma/\sqrt{2})$-subgaussian distribution.

\restylefloat{table}

\begin{table}[h!]
\centering
\begin{tabular}{|c|c|c|c|c|c|c|c|c|c|}
\hline
\multirow{2}{*}{$n$} & \multirow{2}{*}{$\rho$} & \multirow{2}{*}{$\gamma-\beta$} & \multirow{2}{*}{$\alpha$} & \multirow{2}{*}{$k$} & \multirow{2}{*}{$\eta$} & \multicolumn{2}{c|}{$\norm{\tilde{\textbf{s}} - \textbf{s}}_1$} & 
    \multicolumn{2}{c|}{Rejection Rate}\\
\cline{7-10}
& & & & & & $m=10^4$ & $m=10^5$ & $m=10^4$ & $m=10^5$\\
\hline
100 & 39 & 256 & 29 & 1 & 1 & 0 & 0 & 52.09\% & 52.59\%\\
100 & 39 & 256 & 26 & 2 & 1 & 0 & 0 & 52.09\% & 52.02\% \\
100 & 39 & 256 & 25 & 3 & 1 & 0 & 0 & 54.81\% & 55.75\% \\
100 & 39 & 512 & 57 & 1 & 1 & 13 & 0 & 46.13\% & 46.06\% \\
100 & 39 & 512 & 54 & 2 & 1 & 30 & 17 & 58.58\% & 59.52\% \\
100 & 39 & 512 & 50 & 3 & 1 & 13 & 12 & 51.74\% & 51.68\% \\
256 & 75 & 512 & 36 & 1 & 1 & 2 & 0 & 46.36\% & 46.67\% \\
256 & 75 & 512 & 33 & 2 & 1 & 5 & 0 & 43.68\% & 43.88\% \\
256 & 75 & 512 & 32 & 3 & 1 & 2 & 0 & 47.69\% & 47.84\% \\
\hline
\end{tabular}
    \caption{Simulation Results with Subgaussian \textbf{y}}
    \label{tab:securityparameters1}
\end{table}

\begin{table}[h!]
\centering
\begin{tabular}{|c|c|c|c|c|c|c|c|c|}
\hline
\multirow{2}{*}{$n$} & \multirow{2}{*}{$\rho$} & \multirow{2}{*}{$\gamma-\beta$} & \multirow{2}{*}{$k$} & \multirow{2}{*}{$\eta$} & \multicolumn{2}{c|}{$\norm{\tilde{\textbf{s}} - \textbf{s}}_1$} & 
    \multicolumn{2}{c|}{Rejection Rate}\\
\cline{6-9}
& & & & & $m=10^4$ & $m=10^5$ & $m=10^4$ & $m=10^5$\\
\hline
100 & 39 & 256 & 1 & 1 & 47 & 52 & 55.28\% & 57.14\% \\
100 & 39 & 256 & 2 & 1 & 44 & 44 & 47.95\% & 49.24\%\\
100 & 39 & 256 & 3 & 1 & 49 & 46 & 51.33\% & 52.68\%\\
100 & 39 & 512 & 1 & 1 & 66 & 63 & 53.92\% & 53.63\%\\
100 & 39 & 512 & 2 & 1 & 104 & 90 & 49.46\% & 49.26\%\\
100 & 39 & 512 & 3 & 1 & 129 & 131 & 48.47\% & 47.14\%\\
256 & 75 & 512 & 1 & 1 & 71 & 59 & 47.06\% & 45.85\%\\
256 & 75 & 512 & 2 & 1 & 115 & 68 & 49.90\% & 50.64\%\\
256 & 75 & 512 & 3 & 1 & 135 & 49 & 48.32\% & 47.62\%\\
\hline
\end{tabular}
    \caption{Simulation Results with Uniform Random \textbf{y} and Small $\gamma-\beta$}
    \label{tab:securityparameters2}
\end{table}

\begin{table}[h!]
\centering
\begin{tabular}{|c|c|c|c|c|c|c|c|c|}
\hline
\multirow{2}{*}{$n$} & \multirow{2}{*}{$\rho$} & \multirow{2}{*}{$\gamma-\beta$} & \multirow{2}{*}{$k$} & \multirow{2}{*}{$\eta$} & \multicolumn{2}{c|}{$\norm{\tilde{\textbf{s}} - \textbf{s}}_1$} & %
    \multicolumn{2}{c|}{Rejection Rate}\\
\cline{6-9}
& & & & & $m=10^4$ & $m=10^5$ & $m=10^4$ & $m=10^5$\\
\hline
100 & 39 & 4096 & 3 & 1 & 842 & 318 & 57.24\% & 57.06\% \\
100 & 39 & 4096 & 3 & 2 & 940 & 475 & 57.08\% & 57.10\% \\
100 & 39 & 4096 & 3 & 4 & 1039 & 638 & 59.08\% & 58.68\% \\
100 & 39 & 8192 & 3 & 1 & 1850 & 548 & 54.56\% & 55.42\% \\
100 & 39 & 8192 & 3 & 2 & 2014 & 697 & 54.99\% & 55.52\% \\
100 & 39 & 8192 & 3 & 4 & 1975 & 886 & 55.32\% & 55.36\% \\
256 & 75 & 4096 & 3 & 1 & 1753 & 661 & 52.78\% & 52.61\% \\
256 & 75 & 4096 & 3 & 2 & 1873 & 790 & 60.67\% & 60.87\% \\
256 & 75 & 4096 & 3 & 4 & 1972 & 1112 & 75.25\% & 75.08\% \\
\hline
\end{tabular}
    \caption{Simulation Results with Uniform Random \textbf{y} and Large $\gamma-\beta$}
    \label{tab:securityparameters3}
\end{table}
 
Let us make a few initial observations. The first results to take note of are specifically from Table \ref{tab:securityparameters1}. Though these parameters are much smaller than what would be used in practice, it is clear the least squares attack is quite successful in exact recovery of $\textbf{s}$ when a more general subgaussian distribution is used (i.e., not uniform).

The results in Tables \ref{tab:securityparameters2} and \ref{tab:securityparameters3} show that when a uniform distribution is used, the performance of the LSM is far worse. In Table \ref{tab:securityparameters3} the recovered secret gives us a 1-norm which is likely worse than simply guessing a random secret. For chosen $n$ and $k$, there are a total of $n\cdot k$ slots which we must try to determine. Furthermore, we know that are $\rho \cdot k$ nonzero slots. For instance, in the case of $n=100$, $k=3$, $\rho=39$, $\eta=1$, and $\gamma - \beta=4096$, there are 300 total slots of the secret \textbf{s} with $117$ of them being $\pm 1$ and the rest $0$. By randomly choosing a secret $\hat{\textbf{s}}$ with 117 entries $\pm$ 1 we are already guaranteed to have $\norm{\tilde{\textbf{s}} - \textbf{s}}_{1} \leq 417$, with the value of $\norm{\tilde{\textbf{s}} - \textbf{s}}_{1}$ likely being much less than 417. The result in Table \ref{tab:securityparameters3} is not much better, indicating that these parameter choices are secure against the tested sample sizes. This is true for all parameter choices in Table \ref{tab:securityparameters3}. When designing more robust, practical parameter choices, more consideration with larger sample sizes should be considered.

A quite interesting observation from Table \ref{tab:securityparameters2} is the size of $\norm{\tilde{\textbf{s}} - \textbf{s}}_1$ when using $10^4$ samples versus $10^5$ samples. One would assume more samples accessed would result in recovered key which is closer to the true key. Although this is true most of the time, it is not always the case. In fact, 3 out of the 9 parameter instances recovered a key using $10^4$ samples which was at least as good as the recovered key when using $10^5$ samples for the same parameters. These observations point to evidence that, when using rejection sampling and formulating an \textsf{ILWE} instance in this way, taking larger and larger sample sizes may not necessarily result in a better recovered key.

\subsubsection{Remarks on Practical Performance of SVD in Key Recovery}

We remark that although we implemented the attack by Gao outlined in Section \ref{section.svd}, we exclude the experimental results from this paper. For parameter choices similar to those in Tables \ref{tab:securityparameters1} - \ref{tab:securityparameters3}, the SVD attack did not come close to recovering the secret key for any number of samples. In fact, the value of $\norm{\tilde{\textbf{s}} - \textbf{s}}_1$ seemed to be large inconsistent across the number of samples used. For instance, with uniform random $\textbf{y}$, $n=100$, $\rho=39$, $\gamma-\beta=256$, $k=1$, and $\eta=1$ the best value of $\norm{\tilde{\textbf{s}} - \textbf{s}}_1$ with SVD was 139453 with $10^2$ samples. The worst result with SVD for these same parameters was $\norm{\tilde{\textbf{s}} - \textbf{s}}_1 =  6144515$ with $10^4$ samples. 

We should note that these performance results do not necessarily mean the SVD attack does not work or is not effective in a more general case. In fact, when the coefficients of $c\textbf{s}$ are much larger than the coefficients of \textbf{y} and rejection sampling is not used, the SVD attack needs much smaller number of samples than the LSM attack, and the success in full key recovery of the SVD attack is actually quite comparable to the performance of the attack from Section \ref{section.LS_attack}. We omit the full results and discussion on the performance SVD due to the scope of this paper.

\section{Applications and Broader Impact}\label{section.applications}

The findings of this study extend beyond theoretical cryptanalysis and hold practical significance for multiple domains transitioning toward quantum-safe security infrastructures. As quantum computing continues to evolve, industries that depend on the confidentiality and authenticity of digital communication need to reassess the robustness of their cryptographic foundations. The insights derived from our analysis directly contribute to the information for this migration by reinforcing security assurances against recently developed digital attacks.

In the transportation domain, lattice-based digital signature schemes such as CRYSTALS-Dilithium are anticipated to serve as the foundation for authentication mechanisms across Intelligent Transportation Systems (ITS), autonomous and connected vehicles, and Vehicle-to-Everything (V2X) communication infrastructures. Ensuring secure message signing and verification among vehicles, roadside units, and cloud servers is essential to prevent message tampering in safety-critical environments. The resilience insights provided in this work can therefore help to support the design of secure communication protocols emphasizing low latency and small computational costs, which are important requirements for use in transportation infrastructures.

Beyond transportation, the implications of this research extend to a variety of critical infrastructure domains. In finance and digital banking, post-quantum digital signatures will safeguard transaction verification and blockchain-based systems against quantum-enabled forgeries. In healthcare, lattice-based schemes can protect patient telemetry and secure communication between medical devices and cloud servers where message authenticity is vital. The industrial and energy domains, which increasingly depend on distributed sensor networks and remote-control systems, can also benefit from our findings to strengthen authentication between edge devices and supervisory systems. Similarly, defense and aerospace applications, which demand long-term data confidentiality and secure command chains, can leverage \textsf{ILWE}-informed parameterization to anticipate and mitigate quantum-era threats.

Collectively, this work bridges the gap between mathematical cryptanalysis and real-world deployment, emphasizing how subtle variations in lattice-based signature constructions and parameter selections influence resiliency across security-critical domains. The results underscore the growing importance of post-quantum readiness as a cross-domain priority, guiding policymakers, engineers, and system designers toward cryptographic implementations that remain secure in the coming quantum era. Furthermore, this work will be important to different domain areas in terms of how post-quantum signature schemes, an integral part of secure post-quantum cryptography (PQC) communication, may minimize the computational and communication latency required for time-sensitive real-world applications.

\section{Conclusions}\label{section.conclusion}

In this paper, we've conducted an analysis of the effectiveness of the attack from Bootle et al. \cite{bootle} in the context of rejection sampling procedures present in digital signature schemes such as CRYSTALS-Dilithium. In particular, we approach the key recovery problem as an instance of \textsf{ILWE} without any additional information obtained from side-channel attacks while using matrices in $\mathbb{R}$ for representations of constructed polynomials in $\mathcal{R}$. Additionally, we introduce strategies for handling large sample sizes with both the least squares and SVD approaches. Our experimental results show ranges of parameters which are vulnerable to the least squares attack. The effectiveness of the attack when the underlying sampling distribution is changed emphasizes the importance of the standard distributions used. We note that, as expected, the attack is not effective for any parameters used in practice. Additionally, we outline and discuss the relevance of this work and \textsf{ILWE}-based digital signature constructions for practical deployment in critical transportation areas such as ITS and V2X communication. For future work, we plan to pair our simulation results with results for other strategies targeting $\textsf{ILWE}$, using this to develop small, robust ranges of parameters for digital signature schemes utilizing rejection sampling which may reduce communication latency and computational costs needed for use in ITS and other practical applications.

\section*{Funding}

This work is based upon the work supported by the National Center for Transportation Cybersecurity and Resiliency (TraCR) (a U.S. Department of Transportation National University Transportation Center) headquartered at Clemson University, Clemson, South Carolina, USA. Any opinions, findings, conclusions, and recommendations expressed in this material are those of the author(s) and do not necessarily reflect the views of TraCR, and the U.S. Government assumes no liability for the contents or use thereof.

\bibliographystyle{alpha}

\bibliography{bibliography.bib}

@article{crystalsdilithium_ducas_2018,
	title = {{CRYSTALS-Dilithium}: A Lattice-Based Digital Signature Scheme},
	doi = {10.13154/tches.v2018.i1.238-268},
	author = {Ducas, Léo and Kiltz, Eike and Lepoint, Tancrède and Lyubashevsky, Vadim and Schwabe, Peter and Seiler, Gregor and Stehlé, Damien},
	journal = {IACR Transactions on Cryptographic Hardware and Embedded Systems},
	year = {2018},
	litmapsId = {31150964}
}

@techreport{NIST:FIPS204,
  author = {{National Institute of Standards and Technology}},
  title     = {Module-Lattice-Based Digital Signature Standard},
  institution = {U.S. Department of Commerce},
  address= {Washington, D.C.},
   number = {Federal Information Processing Standards Publications (FIPS) 204},
   DOI = {10.6028/NIST.FIPS.204},
   year = {2024},
}

@inproceedings{Ajt96,
author = {Ajtai, M.},
title = {Generating hard instances of lattice problems},
year = {1996},
isbn = {0897917855},
publisher = {Association for Computing Machinery},
address = {New York, NY, USA},
url = {https://doi.org/10.1145/237814.237838},
doi = {10.1145/237814.237838},
booktitle = {Proceedings of the Twenty-Eighth Annual ACM Symposium on Theory of Computing},
pages = {99–108},
numpages = {10},
location = {Philadelphia, Pennsylvania, USA},
series = {STOC '96}
}

@inproceedings{Reg05,
author = {Regev, Oded},
title = {On lattices, learning with errors, random linear codes, and cryptography},
year = {2005},
isbn = {1581139608},
publisher = {Association for Computing Machinery},
address = {New York, NY, USA},
url = {https://doi.org/10.1145/1060590.1060603},
doi = {10.1145/1060590.1060603},
booktitle = {Proceedings of the Thirty-Seventh Annual ACM Symposium on Theory of Computing},
pages = {84–93},
numpages = {10},
location = {Baltimore, MD, USA},
series = {STOC '05}
}

@InProceedings{bootle,
author="Bootle, Jonathan
and Delaplace, Claire
and Espitau, Thomas
and Fouque, Pierre-Alain
and Tibouchi, Mehdi",
editor="Peyrin, Thomas
and Galbraith, Steven",
title="{LWE} Without Modular Reduction and Improved Side-Channel Attacks Against BLISS",
booktitle="Advances in Cryptology -- ASIACRYPT 2018",
year="2018",
publisher="Springer International Publishing",
address="Cham",
pages="494--524",
isbn="978-3-030-03326-2"
}

@InProceedings{cryptoeprint:2013/383,
author="Ducas, L{\'e}o
and Durmus, Alain
and Lepoint, Tancr{\`e}de
and Lyubashevsky, Vadim",
editor="Canetti, Ran
and Garay, Juan A.",
title="Lattice Signatures and Bimodal Gaussians",
booktitle="Advances in Cryptology -- CRYPTO 2013",
year="2013",
publisher="Springer Berlin Heidelberg",
address="Berlin, Heidelberg",
pages="40--56",
isbn="978-3-642-40041-4"
}

@ARTICLE{cryptoeprint:2024/585,
  author={Satriawan, Ardianto and Syafalni, Infall and Mareta, Rella and Anshori, Isa and Shalannanda, Wervyan and Barra, Aleams},
  journal={IEEE Access}, 
  title={Conceptual Review on Number Theoretic Transform and Comprehensive Review on Its Implementations}, 
  year={2023},
  volume={11},
  number={},
  pages={70288-70316},
  doi={10.1109/ACCESS.2023.3294446}}

@misc{gao2025boundeddistancedecodingrandom,
      title={Bounded Distance Decoding for Random Lattices}, 
      author={Shuhong Gao},
      year={2025},
      eprint={2506.16662},
      archivePrefix={arXiv},
      primaryClass={cs.CC},
      url={https://arxiv.org/abs/2506.16662}, 
howpublished = {
    Preprint (arXiv:2506.16662)}
}

@techreport{NIST:FIPS186-5,
  author = {{National Institute of Standards and Technology}},
  title     = {Digital Signature Standard},
  institution = {U.S. Department of Commerce}, 
   address= {Washington, D.C.},
   number = {Federal Information Processing Standards Publications (FIPS) 186-5},
   DOI = {https://doi.org/10.6028/NIST.FIPS.186-5},
   year = {2023},
}

@misc{rfc8032,
    series =    {Request for Comments},
    number =    8032,
    howpublished =  {RFC 8032},
    publisher = {RFC Editor},
    doi =       {10.17487/RFC8032},
    url =       {https://www.rfc-editor.org/info/rfc8032},
    author =    {Simon Josefsson and Ilari Liusvaara},
    title =     {{Edwards-Curve Digital Signature Algorithm (EdDSA)}},
    pagetotal = 60,
    year =      2017,
    month =     jan,
}

@misc{rfc8017,
    series =    {Request for Comments},
    number =    8017,
    howpublished =  {RFC 8017},
    publisher = {RFC Editor},
    doi =       {10.17487/RFC8017},
    url =       {https://www.rfc-editor.org/info/rfc8017},
    author =    {Kathleen Moriarty and Burt Kaliski and Jakob Jonsson and Andreas Rusch},
    title =     {{PKCS \#1: RSA Cryptography Specifications Version 2.2}},
    pagetotal = 78,
    year =      2016,
    month =     nov,
}

@misc{rfc6979,
    series =    {Request for Comments},
    number =    6979,
    howpublished =  {RFC 6979},
    publisher = {RFC Editor},
    doi =       {10.17487/RFC6979},
    url =       {https://www.rfc-editor.org/info/rfc6979},
    author =    {Thomas Pornin},
    title =     {{Deterministic Usage of the Digital Signature Algorithm (DSA) and Elliptic Curve Digital Signature Algorithm (ECDSA)}},
    pagetotal = 79,
    year =      2013,
    month =     aug,
}

@techreport{NIST:SP800-186,
author = {Chen, Lily and Moody, Dustin and  Regenscheid, Andrew and  Robinson, Angela and  Randall, Karen},
  title     = {Recommendations for Discrete Logarithm-based Cryptography: Elliptic Curve Domain Parameters},
   institution = {National Institute of Standards and Technology},
   address= {Gaithersburg, MD},
   number = {NIST Special Publication (SP)  800-186},
   DOI = {https://doi.org/10.6028/NIST.SP.800-186},
   year = {2023}
}

@misc{SPHINCS+,
    howpublished =  {Submission to
the NIST post-quantum project, v.3.1.},
    publisher = {},
    doi =       {},
    url =       {https://sphincs.org/data/sphincs+-r3.1-specification.pdf},
    author =    {Aumasson, Jean-Philippe and Bernstein, Daniel J.  and Beullens, Ward and Dobraunig, Christoph and Eichlseder, Maria and Fluhrer, Scott and Gazdag, Stefan-Lukas and Hülsing, Andreas and Kampanakis, Panos  and Kölbl, Stefan and  Lange, Tanja and
Lauridsen, Martin M.  and Mendel, Florian and Niederhagen, Ruben and Rechberger, Christian and 
Rijneveld, Joost  and Schwabe, Peter  and Westerbaan, Bas},
    title =     {{SPHINCS}+
},
year = 2022
}

@techreport{NIST:FIPS205,
author = {{National Institute of Standards and Technology}},
  title     = {Stateless Hash-Based Digital Signature
Standard},
  institution = {U.S. Department of Commerce}, 
   address= {Washington, D.C.},
   number = {Federal Information Processing Standards Publications (FIPS) 205},
   DOI = {https://doi.org/10.6028/NIST.FIPS.205},
   year = {2024},
}

@misc{FALCON,
    howpublished =  {Submission to
the NIST post-quantum project, Specification v1.2},
    publisher = {},
    doi =       {},
    url =       {https://falcon-sign.info/falcon.pdf},
    author =    
{Prest, Thomas and 
Fouque, Pierre-Alain and
 Hoffstein, Jeffrey and
 Kirchner, Paul and
Lyubashevsky, Vadim  and
 Pornin, Thomas and
 Ricosset, Thomas and
Seiler, Gregor  and
Whyte, William  and
Zhang, Zhenfei },
    title =     {FALCON
},
year = 2020
}

@InProceedings{cryptoeprint:2022/106,
author="Ulitzsch, Vincent Quentin
and Marzougui, Soundes
and Tibouchi, Mehdi
and Seifert, Jean-Pierre",
editor="Smith, Benjamin
and Wu, Huapeng",
title="Profiling Side-Channel Attacks on {Dilithium}",
booktitle="Selected Areas in Cryptography",
year="2024",
publisher="Springer International Publishing",
address="Cham",
pages="3--32",
isbn="978-3-031-58411-4"
}

@article{cryptoeprint:2019/715,
author = {Liu, Yuejun and Zhou, Yongbin and Sun, Shuo and Wang, Tianyu and Zhang, Rui and Ming, Jingdian},
title = {On the Security of Lattice-Based {Fiat-Shamir} Signatures in the Presence of Randomness Leakage},
year = {2021},
issue_date = {2021},
publisher = {IEEE Press},
volume = {16},
issn = {1556-6013},
url = {https://doi.org/10.1109/TIFS.2020.3045904},
doi = {10.1109/TIFS.2020.3045904},
journal = {IEEE Transactions on Information Forensics and Security},
month = jan,
pages = {1868–1879},
numpages = {12}
}

@article{cryptoeprint:2023/896,
author = {Coron, Jean-Sébastien and Gérard, François and Trannoy, Matthias and Zeitoun, Rina},
year = {2023},
month = {08},
pages = {110-145},
title = {Improved Gadgets for the High-Order Masking of {Dilithium}},
volume = {2023},
journal = {IACR Transactions on Cryptographic Hardware and Embedded Systems},
doi = {10.46586/tches.v2023.i4.110-145}
}

@article{shordetailedpaper,
author = {Shor, Peter W.},
title = {Polynomial-Time Algorithms for Prime Factorization and Discrete Logarithms on a Quantum Computer},
journal = {SIAM Journal on Computing},
volume = {26},
number = {5},
pages = {1484-1509},
year = {1997},
doi = {10.1137/S0097539795293172},

URL = { ttps://doi.org/10.1137/S0097539795293172},
eprint = {https://doi.org/10.1137/S0097539795293172}
}

@INPROCEEDINGS{shororiginalpaper,
  author={Shor, P.W.},
  booktitle={Proceedings 35th Annual Symposium on Foundations of Computer Science}, 
  title={Algorithms for quantum computation: discrete logarithms and factoring}, 
  year={1994},
  volume={},
  number={},
  pages={124-134},
  doi={10.1109/SFCS.1994.365700}}

@inproceedings{groverorigpaper,
author = {Grover, Lov K.},
title = {A fast quantum mechanical algorithm for database search},
year = {1996},
isbn = {0897917855},
publisher = {Association for Computing Machinery},
address = {New York, NY, USA},
url = {https://doi.org/10.1145/237814.237866},
doi = {10.1145/237814.237866},
booktitle = {Proceedings of the Twenty-Eighth Annual ACM Symposium on Theory of Computing},
pages = {212–219},
numpages = {8},
location = {Philadelphia, Pennsylvania, USA},
series = {STOC '96}
}

@article{groverexpandedpaper,
  title = {Quantum Mechanics Helps in Searching for a Needle in a Haystack},
  author = {Grover, Lov K.},
  journal = {Physical Review Letters},
  volume = {79},
  issue = {2},
  pages = {325--328},
  numpages = {0},
  year = {1997},
  month = {July},
  publisher = {American Physical Society},
  doi = {10.1103/PhysRevLett.79.325},
  url = {https://link.aps.org/doi/10.1103/PhysRevLett.79.325}
}

@article{cryptoeprint:2023/050,
author = {Berzati, Alexandre and Viera, Andersson and Chartouny, Maya and Madec, Steven and Vergnaud, Damien and Vigilant, David},
year = {2023},
month = {08},
pages = {188-210},
title = {Exploiting Intermediate Value Leakage in {Dilithium}: A Template-Based Approach},
volume = {2023},
journal = {IACR Transactions on Cryptographic Hardware and Embedded Systems},
doi = {10.46586/tches.v2023.i4.188-210}
}

@InProceedings{cryptoeprint:2024/1373,
author="Azevedo-Oliveira, Paco
and Viera, Andersson Calle
and Cogliati, Beno{\^i}t
and Goubin, Louis",
editor="Tauman Kalai, Yael
and Kamara, Seny F.",
title="Uncompressing {Dilithium}'s Public Key",
booktitle="Advances in Cryptology -- CRYPTO 2025",
year="2025",
publisher="Springer Nature Switzerland",
address="Cham",
pages="417--443",
isbn="978-3-032-01855-7"
}

@InProceedings{cryptoeprint:2025/820,
author="Damm, Simon
and Kraus, Nicolai
and May, Alexander
and Nowakowski, Julian
and Thietke, Jonas",
editor="Jager, Tibor
and Pan, Jiaxin",
title="One Bit to Rule Them All -- Imperfect Randomness Harms Lattice Signatures",
booktitle="Public-Key Cryptography -- PKC 2025",
year="2025",
publisher="Springer Nature Switzerland",
address="Cham",
pages="284--316",
isbn="978-3-031-91820-9"
}

@misc{c54b603f579b48a08b698bde47b71455,
    author       = {Rachel Player},
    title        = {Parameter selection in lattice-based cryptography ({PhD} thesis)},
year = {2018},
    howpublished = {Royal Holloway, University of London}
}

@article{vonNeumann1951,
  author  = {John von Neumann},
  title   = {Various techniques used in connection with random digits},
  journal = {Journal of Research of the National Bureau of Standards, Applied Mathematics Series},
  volume  = {12},
  pages   = {36--38},
  year    = {1951}
}

@InProceedings{cryptoeprint:2013/069,
author="Micciancio, Daniele
and Peikert, Chris",
editor="Canetti, Ran
and Garay, Juan A.",
title="Hardness of {SIS} and {LWE} with Small Parameters",
booktitle="Advances in Cryptology -- CRYPTO 2013",
year="2013",
publisher="Springer Berlin Heidelberg",
address="Berlin, Heidelberg",
pages="21--39",
isbn="978-3-642-40041-4"
}

@inproceedings{cryptoeprint:2008/481,
author = {Peikert, Chris},
title = {Public-key cryptosystems from the worst-case shortest vector problem},
year = {2009},
isbn = {9781605585062},
publisher = {Association for Computing Machinery},
address = {New York, NY, USA},
url = {https://doi.org/10.1145/1536414.1536461},
doi = {10.1145/1536414.1536461},
booktitle = {Proceedings of the Forty-First Annual ACM Symposium on Theory of Computing},
pages = {333–342},
numpages = {10},
location = {Bethesda, MD, USA},
series = {STOC '09}
}

@INPROCEEDINGS{5497885,
  author={Regev, Oded},
  booktitle={2010 IEEE 25th Annual Conference on Computational Complexity}, 
  title={The Learning with Errors Problem (Invited Survey)}, 
  year={2010},
  volume={},
  number={},
  pages={191-204},
  doi={10.1109/CCC.2010.26}}

@INPROCEEDINGS{1181960,
  author={Micciancio, Daniele},
  booktitle={The 43rd Annual IEEE Symposium on Foundations of Computer Science, 2002. Proceedings.}, 
  title={Generalized compact knapsacks, cyclic lattices, and efficient one-way functions from worst-case complexity assumptions}, 
  year={2002},
  volume={},
  number={},
  pages={356-365},
  doi={10.1109/SFCS.2002.1181960}}

@article{10.1561/0400000074,
author = {Peikert, Chris},
title = {{A Decade of Lattice Cryptography}},
year = {2016},
issue_date = {Mar 2016},
publisher = {Now Publishers Inc.},
address = {Hanover, MA, USA},
volume = {10},
number = {4},
issn = {1551-305X},
url = {https://doi.org/10.1561/0400000074},
doi = {10.1561/0400000074},
journal = {Foundations and Trends in Theoretical Computer Science},
month = mar,
pages = {283–424},
numpages = {145}
}

\pagebreak

\end{document}